# Mining and Analyzing Twitter trends: Frequency based ranking of descriptive Tweets

Ranking of tweets based on high frequency words

Rishabh Jain
Cisco Systems India Pvt. Ltd.
Bengaluru, India
rishajai@cisco.com

Abhishek B.S.
Cisco Systems India Pvt. Ltd.
Bengaluru, India
abhbs@cisco.com

Satvik Jagannath
Cisco Systems India Pvt. Ltd.
Bengaluru, India
satjagan@cisco.com

*Abstract*—One of the major sources of trending news, events and opinion in the current age is micro blogging. Twitter, being one of them, is extensively used to mine data about public responses and event updates. This paper intends to propose methods to filter tweets to obtain the most accurately descriptive tweets, which communicates the content of the trend. It also potentially ranks the tweets according to relevance. The principle behind the ranking mechanism would be the assumed tendencies in the natural language used by the users. The mapping frequencies of occurrence of words and related hash tags is used to create a weighted score for each tweet in the sample space obtained from twitter on a particular trend.

## I. INTRODUCTION

Trend Analysis is spotting a pattern of occurrence or tendencies that data or information can portray. Trend analysis is often used to predict future events, like the likeability of a product, or the viewership of a match, etc. Micro blogging is a medium of broadcast which is similar to blogging. The aggregate and the actual blog size are typically smaller than a traditional blog. Micro blogs allow users to share content such as short sentences, individual images or video links.

Twitter is one of the most popular micro blogging websites on the Internet today. It uses 'Hashtags', which is Hash '#', followed by the trend-name, to identify and group micro-posts about a particular trend. Micro-posts in twitter are known as 'tweets'. Twitter also categorizes tweets and trends according to region and on its web interface it displays the most recent tweets as feeds. Twitter also provides handles in the form of API for developers to create applications centered on the data gathered by tweets and trends.

The data gathered by the tweets is generally unprocessed and to determine any significant conclusions from it requires it to be mined and processed. The Hashtags generally are telegram style word(s) description without spaces and attempts to summarize the trend, but seldom is a trend completely understood by its hash tag. The methods proposed in this paper takes sample data from twitter for the most trending Hashtags and attempts to summarize the trend by displaying the most descriptive tweets which can summarize the content of the trend. It also prescribes a weighted score to the sampled tweets in order to rank it according to accuracy of its description in its content about the Hashtag. This would enable a better informing feed about the trend than a 'recent' based display currently used in the web interface.

## II. ASSUMPTIONS AND SCOPE OF WORK

It is assumed that the tweets in themselves are not highly descriptive. One of the constraints that Twitter allows a maximum of 140 characters, alphanumeric and special characters that can be used in a tweet. This constraint makes tweets be in a telegram format where key words are prioritized over grammatical correctness.

The scope of the methods proposed in this paper extends to assign scores to the tweets taken as sample space so as to be able to rank them as highest descriptive with the highest score. This can be used to accustom a new user to get acquainted with the trend content.

## III. DESCRIPTION OF THE ALGORITHM

The algorithm takes as input a sample space of 'n' predefined number of tweets. It also takes the highest trending 'x' number of trends. The output of running the algorithm is the tweets of the sample space ranked with a deceasing description index.

The algorithm also uses two dictionaries. The first dictionary contains the list of the words which have less significance to the content description and are more grammatical tools, namely articles, prepositions and conjunctions. The second dictionary consists of all common nouns, adjectives, adverbs, verbs and their derivatives. We will call the former 'filter' and the latter 'cnfilter'.

We use the sample space in a file, separated by an end of tweet character, like '%%'. Once the tweets are acquired, we find the frequency of every word that is used in the file containing the tweet sample space. This would exclude the '#' tags and the '@' tags. We also ignore the URLs in the tweets while finding the frequencies. Hence the list of words and their corresponding frequencies is prepared and stored.

We now check for association of the highest trending tweet with the other high trending tweets. Tweets about the same event, or person, hold useful content and can be assumed to contain more relevant data. We use the tweets with a high trending hashtag along with the highest trending hashtag for the second time to collect the frequency to update the previously generated frequency table.

Once the frequency list is obtained we perform a rating on the words to find its weighted score. This weighted score is used to get the cumulative score of each tweet which can be used to rank the tweets according to its content relevance.

We also propose a way to learn from the newer tweets about the hashtag and get more accurate tweet ranking.

## IV. SAMPLE SPACE PREPARATION

The first requirement is to find a sample space for the algorithm to run. This finite sample space is a subset of the data that is available at twitter. For the purpose of the explanation, the number of tweets taken in the sample space 'SS' is 'n'. This number 'n' when increased will improve the accuracy of the analysis but also increase the average runtime, but when decreased it radically decreases the accuracy of the findings.

When writing the SS into a file, we have to make sure of that they are logically separated by either an end of tweet character or the file format is such that individual tweets can be logically separated.

As is used in the algorithm, we also store the top trending 'x' hashtags along with the SS.

```
For tweet in retrived_tweets
  If tweet has highest_tag
    Add tweet to SS
    // Adding again into SS
    For tag in x
      If tweet has tag
        Add tweet to SS
```

## V. TWEET RANKING ALGORITHM

The Algorithm is run in two stages. The first stage is the word frequency index and the second stage is tweet rating.

### A. Frequency Index

The Frequency of the words that are used in the SS is used as the fundamental data for content rating. The Algorithm for calculating frequency index is as follows:

```
Word_count = 0
For word in SS
  If word in Freq_list
    Freq_list[word] = Freq_list[word] + 1
  Else
    Freq_list[word] = 1
  Word_count = word_count + 1
For word in Freq_list
  Freq_index[word] = Freq_list[word] / Word_count
```

### B. Tweet Score

The Tweet score is an index generated for each Tweet 'T' in the sample space 'SS' which will be the basis for the rankings. The generated rankings would be in the decreasing order of the Tweet score.

The Algorithm of for the generation of the Tweet score is as follows:

```
For Tweet in SS
  Score[Tweet] = 0
  For word in Tweet
    If word not in filter
      Score[Tweet] = Score[Tweet] + Freq_index[word]
    If word in cnfilter
      Cnindex[Tweet] = Cnindex[Tweet] + 1
  If Cnindex[Tweet] = word_count(Tweet)
    allCn[Tweet] = True
  Else allCn[Tweet] = False
```

### C. Dynamic Learning

While we are generating the score for each tweet in the sample space, we are overlooking the newly broadcasted tweets made after the sample space collection. Hence, the generated score might become obsolete after a finite time.

To accommodate dynamic learning in the algorithm we introduce a learning index 'LI' which determines the rate of decomposition of the previously collected data and the exclusive inclusion of the newer data. To determine the 'LI' we need to decide the decomposition rate. If we want the contribution of the old data to reduce to 't' percent after 'f' new incoming tweet streams then the value of 'LI' can be calculated as,

$$t = n (LI)^f$$

$$LI = (t/n)^{(1/f)}$$

Once this Learning Index LI is calculated the Dynamic Algorithm can be stated as

```
Word_count = LI * Word_count
For word in new_SS
  If word not in new_Freq_list
    new_Freq_list[word] = (LI * Freq_list[word]) + 1
  Else
    new_Freq_list[word] = new_Freq_list[word] + 1
  Word_count = Word_count + 1
For word in new_Freq_list
  Freq_index[word] = Freq_list[word] / Word_count
```

## VI. GENERATED RESULTS

Since the frequency of words is the primary source of classification in this ranking algorithm, it is important to observe the distinguishing frequency. For the target high frequency words to contribute more towards the score of a Tweet its frequency should ideally be substantially more than the more common words, whose contribution should be

insignificant. We can observe the trend in a sample run of the frequency index algorithm as follows

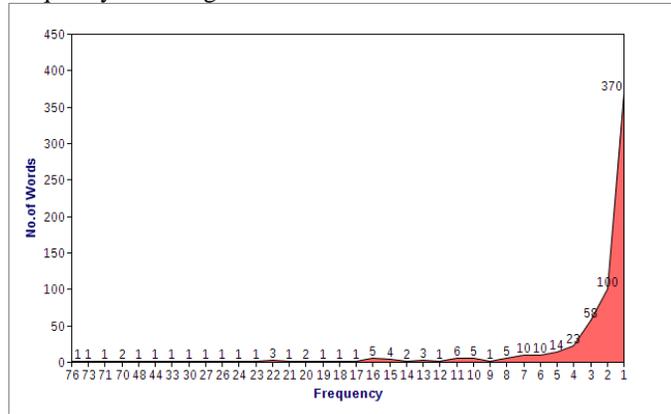

Figure 1. Graph of No. of Words vs. their frequency

As we can see, the 22 highest frequency words, among 549 words contribute approximately 56% of the Sample Space SS content. This shows the tendency of keyword based algorithms to have a definite validity.

## VII. EXAMPLE

On running the above algorithm, we get the following results,

| Highest Trending : #BollywoodMoviesEnglishTitles | |
|---|---|
| **Word** | **Frequency** |
| More | 24 |
| Trending | 23 |
| Stand | 22 |
| Total | 22 |
| Winners | 22 |
| **Most Eligible Tweet :** | |
| So Tweeple, we're trending at No.1!!! 2 more winners! so our contest's total winners now stand at 7!!! | |

In the above run the most eligible, or the most highly ranked tweet, indicates that the trend is a contest where the winners are decided according to the highest trending. We can again see the high frequency word tendency to be different from common word usage.

In the above example the SS contained 500 tweets as a sample space.

## VIII. FUTURE IMPROVEMENTS

In the current implementation we do not consider the length of the tweet. As per restrictions, a tweet may not exceed 140 characters, but when we run the proposed algorithms we do not consider whether the tweet is using the maximum available character space. One way to do it is to calculate the Tweet score in the following manner,

$$Tweet\_score = [\sum word\_score] * (word\_count \div 140)$$

In this way, the Tweet that is most highly ranked is considered to be using the maximum character space available. The results generated after using this enhancement is beyond the scope of this literature.

## CONCLUSION

Considering the results generated by the algorithms produced by this study, the Tweets can successfully be used to describe the content of the trends. Since the ranking algorithms logically hold true for most of the tweet types, it is still not a good way to detect spam or RT, ReTweets. For implementing those features Image and URL checking algorithms needs to be in place. Also these algorithms do not take into considerations the 'favorite' attribute that is associated with each tweet. A 'favorite' attribute is a counter which is incremented whenever a user up votes a particular tweet.

## ACKNOWLEDGEMENT

We would like to thank Rajat Tandon for helping collect data and for providing feedback. We would also thank Sakshi Gupta, Rupanta Rwiteej Dutta, Samyak Baliar Singh and Vinay Menon for their insights and the motivation.

## REFERENCES

[1] boyd, danah, Scott Golder, and Gilad Lotan. 2010. "Tweet, Tweet, Retweet: Conversational Aspects of Retweeting on Twitter." HICSS-43. IEEE: Kauai, HI, January 6

[2] Shirley Ann Williams, Melissa Terras, Claire Warwick (2013). "What people study when they study Twitter: Classifying Twitter related academic papers". Journal of Documentation, 69 (3).

[3] Y.-Y. Ahn, S. Han, H. Kwak, S. Moon, and H. Jeong. Analysis of topological characteristics of huge online social networking services. In Proc. of the 16th international conference on World Wide Web. ACM, 2007.

[4] R. Crane and D. Sornette. Robust dynamic classes revealed by measuring the response function of a social system. Proc. of the National Academy of Sciences, 105(41):15649–15653, 2008.

[5] J. Leskovec, J. Kleinberg, and C. Faloutsos. Graphs over time: densification laws, shrinking diameters and possible explanations. In Proc. of the 11th ACM SIGKDD international conference on Knowledge discovery in data mining. ACM, 2005.

[6] D. Zhao and M. B. Rosson. How and why people twitter: the role that micro-blogging plays in informal communication at work. In Proceedings of the ACM 2009 international conference on Supporting group work. ACM, 2009.

[7] C. Wilson, B. Boe, A. Sala, K. P. Puttaswamy, and B. Y. Zhao. User interactions in social networks and their implications. In Proc. of the 4th ACM European conference on Computer systems. ACM, 2009.

[8] Twitter Search API. http://apiwiki.twitter.com/Twitter-API-Documentation.

[9] J. Leskovec, L. A. Adamic, and B. A. Huberman. The dynamics of viral marketing. In Proc. of the 7th ACM conference on Electronic commerce. ACM, 2006.

[10] [20] J. Leskovec, L. Backstrom, and J. Kleinberg. Meme-tracking and the dynamics of the news cycle. In Proc. of the 15th ACM SIGKDD international conference on Knowledge discovery and data mining. ACM, 2009.

[11] M. E. J. Newman and J. Park. Why social networks are different from other types of networks. Phys. Rev. E, 68(3):036122, Sep 2003.

[12] Jin O., Liu N.N., Zhao K., Yu Y., Yang Q. Transferring topical knowledge from auxiliary long texts for short text clustering 2011 International

[13] Conference on Information and Knowledge Management, Proceedings

[14] M. E. J. Newman and J. Park. Why social networks are different from othertypes of networks. Phys. Rev. E, 68(3):036122, Sep 2003.M. Young, The Technical Writer's Handbook. Mill Valley, CA:UniversityScience,1989.